\documentclass[aps,prl,twocolumn,showpacs,10pt,superscriptaddress,preprintnumbers,nofootinbib]{revtex4-1}
\usepackage{epsfig,amssymb,amsmath,psfrag,epstopdf,color}
\pdfoutput=1
\usepackage{graphicx}
\usepackage[margin=7pt,justification=centerlast]{caption}
\usepackage{subcaption}
 
\usepackage{paralist}

\allowdisplaybreaks

\def\beq{\begin{equation}}
\def\eeq{\end{equation}}
\def\bsp#1\esp{\begin{split}#1\end{split}}
\newcommand{\be}{\begin{equation}}
\newcommand{\ee}{\end{equation}}
\newcommand{\bea}{\begin{eqnarray}}
\newcommand{\eea}{\end{eqnarray}}

\def\cO{{\mathcal O}}

\def\to{\rightarrow}

\def\ksl{\not{\hbox{\kern-2.3pt $k$}}}

\def\spa#1.#2{\left\langle#1\,#2\right\rangle}
\def\spb#1.#2{\left[#1\,#2\right]}
\def\lor#1.#2{\left(#1\,#2\right)}
\def\sand#1.#2.#3{%
\left\langle\smash{#1}{\vphantom1}^{-}\right|{#2}%
\left|\smash{#3}{\vphantom1}^{-}\right\rangle}

\newcommand{\nn}{\nonumber}

\newcommand{\mcdot}{\!\cdot}

\begin{document}

\preprint{SLAC-PUB-17203}

\title{The Energy-Energy Correlation at Next-to-Leading Order in QCD,
Analytically}
\author{Lance~J.~Dixon}
\affiliation{SLAC National Accelerator Laboratory, Stanford
  University, Stanford, CA 94039, USA}
\author{Ming-xing~Luo}
\author{Vladyslav Shtabovenko}
\author{Tong-Zhi~Yang}
\author{Hua~Xing~Zhu}
\affiliation{Zhejiang Institute of Modern Physics, Department of
  Physics, Zhejiang University, Hangzhou, 310027, China}

\begin{abstract}
The energy-energy correlation~(EEC) between two detectors in $e^+e^-$
annihilation was computed analytically at leading order in QCD almost
40 years ago, and numerically at next-to-leading order (NLO) starting
in the 1980s.  We present the first analytical result for the EEC at NLO,
which is remarkably simple, and facilitates analytical study of the
perturbative structure of the EEC. We provide the expansion of EEC in the
collinear and back-to-back regions through to next-to-leading power,
information which should aid resummation in these regions.
\end{abstract}

\maketitle


\noindent\textit{Introduction.}
The energy-energy correlation~(EEC)~\cite{Basham:1978bw}
measures particles detected by two detectors at a fixed angular separation
$\chi$, weighted by the product of the particles' energies.
The EEC is an infrared-safe characterization of hadronic energy flow in
$e^+e^-$ annihilation.  It has been used for precision tests of
quantum chromodynamics~(QCD) and measurement of the strong coupling
constant $\alpha_s$~\cite{Acton:1993zh,Abe:1994mf}. In perturbative QCD,
the EEC is defined by
\begin{align}
  \label{eq:def}
  \frac{d\Sigma}{d\cos\chi} = \; \sum_{i,j} \int  \frac{E_i E_j}{Q^2}
  \delta(\vec{n}_i \mcdot \vec{n}_j - \cos\chi) d\sigma \,,
\end{align}
where $i$ and $j$ run over all the final-state massless partons,
which have four-momenta $p_i^\mu$
and $p_j^\mu$~(including the case $i=j$ at $\chi=0$);
$Q^\mu$ is the total four-momentum of the $e^+e^-$ collision and
$d\sigma$ is the phase-space measure.
The three-vectors $\vec{n}_{i,j}$ point along the spatial components of
$p_{i,j}$.  The definition~\eqref{eq:def} implies the sum rule
\begin{align}
  \label{eq:sumRule}
\frac{1}{\sigma}  \int^1_{-1} d\cos\chi \frac{d\Sigma}{d\cos\chi} = 1 \,,
\end{align}
where $\sigma$ is the total cross section for $e^+e^-$ annihilation to hadrons.

The leading order~(LO) QCD prediction for the EEC has been available since
the 1970s~\cite{Basham:1978bw}:
\begin{eqnarray}
&&\frac{1}{\sigma_0} \frac{d \Sigma}{d\cos\chi} \ =\
  \frac{\alpha_s(\mu)}{2 \pi} \, C_F \, \frac{3 - 2 z}{4 (1 - z) z^5}
\label{eq:LO}\\
&&\null\hskip0.01cm
\times \Bigl[ 3 z (2 - 3 z) + 2 (2 z^2 - 6 z + 3) \log(1-z) \Bigr]
 + \cO(\alpha_s^2) \,,
\nn
\end{eqnarray}
where $\sigma_0$ is the Born cross section for $e^+e^- \to q \bar{q}$,
$C_F$ is the quadratic Casimir in the fundamental representation,
and we have introduced $z = (1 - \cos\chi)/2$.
The cross section is strongly peaked at $\chi = 0$ ($z=0$)
and $\chi = \pi$ ($z=1$), regions that require resummation of
logarithms due to emission of soft and collinear partons.
At intermediate angles, higher-order corrections tend to flatten the
distribution.

The EEC was first computed numerically at next-to-leading order~(NLO) in QCD
by several groups in the 1980s and 1990s, originally leading to
conflicting results.  Different methods were used to handle soft
and collinear singularities from real radiation:
phase-space
slicing~\cite{Schneider:1983iu,Falck:1988gb,Glover:1994vz,Kramer:1996qr}
subtraction methods~\cite{Ali:1982ub,Barreiro:1986si,%
  Richards:1982te,Richards:1983sr,Kunszt:1989km,Glover:1994vz,%
  Catani:1996jh,Catani:1996vz},
or hybrid schemes~\cite{Glover:1994vz,Clay:1995sd,Kramer:1996qr}. 
Accurate numerical NLO results are available from the program \textsc{Event2},
based on dipole subtraction~\cite{Catani:1996jh,Catani:1996vz}.
Quite recently, the EEC has been computed at NNLO in
QCD using the CoLoRFulNNLO local subtraction
method~\cite{DelDuca:2016csb,Tulipant:2017ybb}.

In perturbation theory, the EEC is singular in both the
collinear~($z \to 0$) and back-to-back regions~($z \to 1$),
as can be seen explicitly from Eq.~\eqref{eq:LO}.
The leading-logarithmic collinear behavior can be
obtained from the ``jet calculus''
approach~\cite{Konishi:1978yx,Konishi:1978ax},
in terms of the anomalous dimension matrix of twist-two, spin-three
operators~\cite{Konishi:1978ax,Richards:1983sr}.
Resummation of the EEC in the back-to-back (Sudakov)
region has been performed at next-to-leading-logarithmic~(NLL) and
NNLL accuracy~\cite{Collins:1981uk,Ellis:1983fg,deFlorian:2004mp}.
Quite recently, a factorization formula for the EEC has been derived which
permits its resummation to N$^3$LL~\cite{MoultZhu2018}.
Possible non-perturbative corrections to the EEC
have also been investigated~\cite{Dokshitzer:1999sh}.

In ${\cal N}=4$ super-Yang-Mills theory (SYM), the 
EEC has been computed analytically at NLO in terms of classical
polylogarithms~\cite{Belitsky:2013ofa},
using an approach that bypasses the need for infrared cancellations in
intermediate steps~\cite{Belitsky:2013xxa,Belitsky:2013bja}. 
In the strong-coupling limit and at large $N_c$, the EEC in ${\cal N}=4$ SYM
can be calculated using AdS/CFT duality~\cite{Hofman:2008ar}.

Despite all of this progress, the analytic computation of the EEC at NLO in QCD
has remained an open problem, whose solution is desirable for several reasons.
First, the analytical results can settle any remaining discrepancies
between different numerical methods, and provide a benchmark for future
numerical evaluations.
Second, the analytical results allow extraction of the $\cO(\alpha_s^2)$
asymptotic behavior in the collinear and back-to-back regions,
not just at leading power, but any desired power.
Knowledge of the subleading power corrections can be very helpful for
improving the understanding of resummation at subleading
power~\cite{Laenen:2008ux,Laenen:2010uz,Bonocore:2016awd,%
Moult:2016fqy,Boughezal:2016zws,Feige:2017zci,Moult:2017rpl,%
Moult:2017jsg,Balitsky:2017flc,Beneke:2017ztn}.
Third, no other event-shape variable has been computed analytically at NLO.
Calculationally, the EEC appears to be the simplest such
observable. Knowing it analytically at NLO marks an important step in
the perturbative understanding of event-shape observables, and may pave
the way for an analytic computation at NNLO.
Recently, progress has been made toward computing the EEC at NLO
by linearizing the measurement function~\cite{Gituliar:2017umx}.
In this letter, we present the first fully analytic result for the EEC in
QCD at NLO. 


\noindent\textit{The Calculation.} At LO, calculation of the EEC is
straightforward, because only finite phase-space integrals need to be
evaluated. At NLO, the renormalized virtual corrections
contain explicit infrared (IR) poles, but no singularities
from the boundary of phase space.
We use the analytical one-loop amplitudes~\cite{Ellis:1980wv,Garland:2001tf},
and perform the phase-space integral directly. The real radiative
corrections represent the most complicated part of this calculation,
because the phase-space integrals contain unresolved soft and collinear
IR divergences.  We apply reverse
unitarity~\cite{Anastasiou:2002yz,Anastasiou:2003yy} to write on-shell
delta functions as differences of Feynman propagators with
opposite signs for $i \varepsilon$, which allows
the use of integration-by-parts~(IBP)
equations~\cite{Chetyrkin:1981qh,Tkachov:1981wb} for multi-loop integrals.
The EEC measurement function can be written in the same way,
\begin{align}
  \label{eq:measurement}
  \delta({\cal M}_{ij}(\chi)) = \frac{1}{2 \pi i}
  \left(\frac{1}{{\cal M}_{ij}(\chi) - i \varepsilon}
  - \frac{1}{{\cal M}_{ij}(\chi) + i \varepsilon}\right) \,,
\end{align}
where 
${\cal M}_{ij}(\chi) 
= (p_i \mcdot Q \, p_j \mcdot Q) ( \vec{n}_i \mcdot \vec{n}_j - \cos\chi) 
= (p_i \mcdot Q \, p_j \mcdot Q) (1 - \cos\chi) - p_i \mcdot p_j$.
While the application of reverse unitarity to phase-space integrals
is now quite standard, Eq.~\eqref{eq:measurement} is special in the sense
that ${\cal M}_{ij}(\chi)$ is a non-linear function of Lorentz dot products.
In addition to the usual IBP equations, an extra equation,
\begin{eqnarray}
&& \big[ (1-\cos\chi) (p_i \mcdot Q \, p_j \mcdot Q) - p_i \mcdot p_j \big]
  [\delta({\cal M}_{ij}(\chi))]^k
\nn\\
&=&  [\delta({\cal M}_{ij}(\chi))]^{k-1} \,,
\label{eq:extraRule}
\end{eqnarray}
for $k=1,2,\ldots$, with $[\delta({\cal M}_{ij}(\chi))]^0 \equiv 0$,
has to be added in order to fully reduce the phase-space integrals to
master integrals~(MIs). 

In our calculation, we use \textsc{Qgraf}~\cite{Nogueira:1991ex}
to generate the squared amplitudes for the LO and real NLO terms.
We set all quark masses to zero, and ignore contributions from the top quark,
as well as the (tiny) purely axial-vector contributions in the case
of $e^+e^-$ annihilation via the $Z$ boson.
The color and Dirac algebra is evaluated using
\textsc{Form}~\cite{Vermaseren:2000nd}. The resulting tree-level
matrix elements agree fully with Ref.~\cite{Ellis:1980wv}.
The squared matrix elements for the NLO real corrections,
ignoring the EEC measurement function, can be divided into three
integrand topologies, each consisting of nine Feynman propagators
(one in the numerator).  Since there are four partons in the real NLO final
state, there are $({4 \atop 2}) = 6$ different measured pairs to sum over for the EEC.
Multiplying the 3 inclusive integrand topologies by the 6 pairs of measurement
delta functions gives rise to $18$ separate integral topologies.
We use \textsc{LiteRed}~\cite{Lee:2012cn,Lee:2013mka} to generate the
standard IBP equations for these integral families, and then add the
additional integral relation~\eqref{eq:extraRule} manually.
We then export the resulting IBP relations to
\textsc{Fire}~\cite{Smirnov:2008iw,Smirnov:2014hma} to perform the
integral reduction, which leads to a total of 40 independent MIs.

We solve for the MIs by the method of differential
equations (DEs)~\cite{Kotikov:1990kg,Gehrmann:2000zt},
and convert the DE systems into a canonical
form~\cite{Henn:2013pwa}.  Some of the DE systems can be converted to
canonical form using the original variable $z$; for
others, an algebraic change of variable to $x = \sqrt{z}$ or
$y = i\sqrt{z}/\sqrt{1-z}$ is required.  After identifying the
appropriate variable for each integral family,
the conversion to a canonical basis can be automated by the
\textsc{Mathematica} package
\textsc{Fuchsia}~\cite{Gituliar:2017vzm}.
The resulting symbol alphabet, characterizing the arguments
of the polylogarithms, is $\{ 1 - x \,, y \,, 1-y \,, 1+y \}$.
Note that $z$, $1-z$, $x$ and $1+x$ also appear, but are not
multiplicatively independent, since $1/(1-y^2) = 1-x^2 = 1-z$, etc.,
so we do not count them as separate symbol letters.
This alphabet implies that the solution to the DEs can be written fully
in terms of harmonic polylogarithms~(HPLs)~\cite{Remiddi:1999ew},
which can be manipulated conveniently using the \textsc{Mathematica}
package \textsc{HPL}~\cite{Maitre:2005uu}.  Our final NLO result contains
at most weight 3 HPLs, which can all be reduced to classical polylogarithms.

The most intricate part of the calculation is the determination of the
constants of integration for the DEs, which requires combining several
different constraints.  First, we require that the leading power
expansion $z^\alpha$
of each MI in the collinear limit~$z\to0$ has the correct power
$\alpha$, which can be predicted by simple power counting.
We find that all the MIs in our problem have
at most a $z^{-1}$ pole.  (Some have $z^0$ or $z^1$ as their leading behavior.)
Requiring the absence of $z^{-2}$ or worse poles
strongly constrains the boundary constants.
The second constraint is the $z \to - \infty$ limit:
Before converting to the canonical basis,
MIs that are pure functions of uniform transcendental weight should
vanish in this limit. 
The third constraint comes from performing a weighted integral over $z$,
which allows the removal of the measurement constraint,
according to the integral relation
\begin{align}
  & \int dPS^{(4)} \hat{z}_{ij}^n (1-\hat{z}_{ij})^m  \, {\cal I} (\{ p\})
\nn\\
= \; & \int^1_0 dz \, z^n (1-z)^m
  \int dPS^{(4)} \, {\cal I} (\{p\}) 
\nn\\
& \, \times 2 \, p_i \mcdot Q \, p_j \mcdot Q \, \delta ({\cal M}_{ij} (\chi) ) \,.
  \label{eq:relation}
\end{align}
Here $dPS^{(4)}$ is the four-particle Lorentz-invariant phase-space
measure in $D$ dimensions,
$\hat{z}_{ij} = Q^2 \, p_i \mcdot p_j (2 \, p_i \mcdot Q \, p_j \mcdot Q)^{-1}$,
and ${\cal I}(\{p\})$ denotes a MI integrand.
We choose the integers $n$ and $m$ to be sufficiently positive that
the particular integral over $z$ converges, and $n\leq1$, $m\leq1$ to keep
the IBP reduction tractable.
The integral on the left-hand side can be reduced
to known inclusive four-particle phase-space
integrals~\cite{Gehrmann-DeRidder:2003pne}, if we multiply the integrand
on both sides by $(p_i \mcdot Q \, p_j \mcdot Q)^{n+m}$.
The last constraint we apply is to demand that the full NLO real
corrections, after substituting in the results for the MIs, have at
most a $z^{-1}$ pole.  This gives extra constraints, beyond the
constraints applied to the individual MIs. A similar method has
been applied to fix constants of integration for DEs for
auxiliary EEC MIs~\cite{Gituliar:2017umx}.


\noindent\textit{The result.} After combining the real and virtual
corrections, and adding the counterterm to renormalize $\alpha_s$,
we obtain our final result for the EEC at NLO.
We write the differential distribution as
\begin{align}
  \label{eq:diffdis}
  \frac{1}{\sigma_0} \frac{d\Sigma}{d\cos\chi} = &
 \frac{\alpha_s(\mu)}{2
  \pi} A(z) + \left(\frac{\alpha_s(\mu)}{2 \pi}\right)^2
\nn\\
&
 \times  \left(\beta_0
  \log \frac{\mu}{Q} A(z) + B(z) \right)  + {\cal O}(\alpha_s^3) \,,
\end{align}
where the LO coefficient $A(z)$ has already been given in Eq.~\eqref{eq:LO},
and $\beta_0 = 11C_A/3 -  4 N_f T_f/3$. For QCD with $N_f$ flavors of quarks,
$C_A = N_c = 3$, $C_F = (N_c^2-1)/(2N_c) = 4/3$, and $T_f = 1/2$.
The NLO coefficient $B(z)$
can be further decomposed into different color structures,
\begin{align}
  \label{eq:colordec}
  B = C_F^2 B_{\text{lc}}  + C_F (C_A - 2 C_F) B_{\text{nlc}}
    + C_F N_f T_f B_{N_f}  \,.
\end{align}
We have calculated each coefficient in the color decomposition analytically.
The leading-color correction $B_{\text{lc}}$ reads,
\begin{widetext}
  \begin{align}
 B_{\text{lc}} = \; 
& +\frac{122400 z^7-244800 z^6+157060 z^5-31000 z^4+2064 z^3+72305 z^2-143577 z+63298}{1440 (1-z) z^4}
\nn \\
& -\frac{-244800 z^9+673200 z^8-667280 z^7+283140 z^6-48122 z^5+2716 z^4-6201 z^3+11309 z^2-9329 z+3007}{720 (1-z) z^5} g^{(1)}_1
\nn \\
& -\frac{244800 z^8-550800 z^7+422480 z^6-126900 z^5+13052 z^4-336 z^3+17261 z^2-38295 z+19938}{720 (1-z) z^4} g^{(1)}_2
\nn \\
& +\frac{4 z^7+10 z^6-17 z^5+25 z^4-96 z^3+296 z^2-211 z+87}{24 (1-z) z^5} g^{(2)}_1
\nn \\
& +\frac{-40800 z^8+61200 z^7-28480 z^6+4040 z^5-320 z^4-160 z^3+1126 z^2-4726 z+3323}{120 z^5} g^{(2)}_2
\nn \\
& -\frac{1-11 z}{48 z^{7/2}} g^{(2)}_3
 -\frac{120 z^6+60 z^5+160 z^4-2246 z^3+8812 z^2-10159 z+4193}{120 (1-z) z^5} g^{(2)}_4
\nn \\
& -2 \left(85 z^4-170 z^3+116 z^2-31 z+3\right) g^{(3)}_1
 +\frac{-4 z^3+18 z^2-21 z+5}{6 (1-z) z^5} g^{(3)}_2
 +\frac{z^2+1}{12 (1-z)} g^{(3)}_3 \,,
\label{eq:lc}
  \end{align}
where the $g_m^{(n)}$ are pure functions of uniform transcendental weight $n$.
Their explicit definitions are
\begin{gather}
  g_1^{(1)} = \log (1-z)\,, \qquad g_2^{(1)} =  \log (z)\,,
\qquad
g_1^{(2)} = 2 (\text{Li}_2(z)+\zeta_2)+\log ^2(1-z) \,, 
\nn\\
g_2^{(2)} = \text{Li}_2(1-z)-\text{Li}_2(z) \,, \qquad
g_3^{(2)} = - 2 \, \text{Li}_2\left(-\sqrt{z}\right)
+ 2 \, \text{Li}_2\left(\sqrt{z}\right)
+ \log\left(\frac{1-\sqrt{z}}{1+\sqrt{z}}\right) \log (z) \,,\qquad
g_4^{(2)} = \zeta_2 \,,
\nn\\
g_1^{(3)} = -6
\left[ \text{Li}_3\left(-\frac{z}{1-z}\right)-\zeta_3 \right]
- \log \left(\frac{z}{1-z}\right)
 \left(2 (\text{Li}_2(z)+\zeta_2)+\log^2(1-z)\right)
\,,
\nn\\
g_2^{(3)} = -12
\left[ \text{Li}_3(z)+\text{Li}_3\left(-\frac{z}{1-z}\right) \right]
+ 6 \, \text{Li}_2(z) \log(1-z) + \log^3(1-z) \,,
\nn\\
g_3^{(3)} = 6 \log(1-z) \, (\text{Li}_2(z)-\zeta_2)
- 12 \, \text{Li}_3(z) + \log^3(1-z) \,.
\label{eq:gdef}
\end{gather}
\end{widetext}
Note that $B_{\text{lc}}$ contains explicit dependence on $\sqrt{z}$
through the function $g_3^{(2)}$ and its coefficient, whose product is
even under $\sqrt{z} \to - \sqrt{z}$. This property also holds in
${\cal N}=4$ SYM~\cite{Belitsky:2013ofa}. To describe $B_{\text{lc}}$,
we need just two weight 1, four weight 2, and three weight 3 transcendental
functions.  To express $B_{\text{nlc}}$ and $B_{N_f}$ requires two more
weight 3 transcendental functions.~\footnote{%
The NLO EEC in ${\cal N}=4$ SYM~\cite{Belitsky:2013ofa}, after some
rearrangement, can be expressed in terms of a subset of the
transcendental functions needed for QCD.}
Individual virtual and real contributions
contain HPLs with argument $y = i \sqrt{z}/\sqrt{1-z}$. 
However, they cancel out in the final physical result.
The explicit expressions for $B_{\text{nlc}}$ and $B_{N_f}$ can be found in
the supplemental material for this letter.
In an ancillary file, we provide computer-readable expressions for all
these functions, as well as their behavior in various limits.

\begin{figure}[ht]
\centering
  \includegraphics[width=\linewidth]{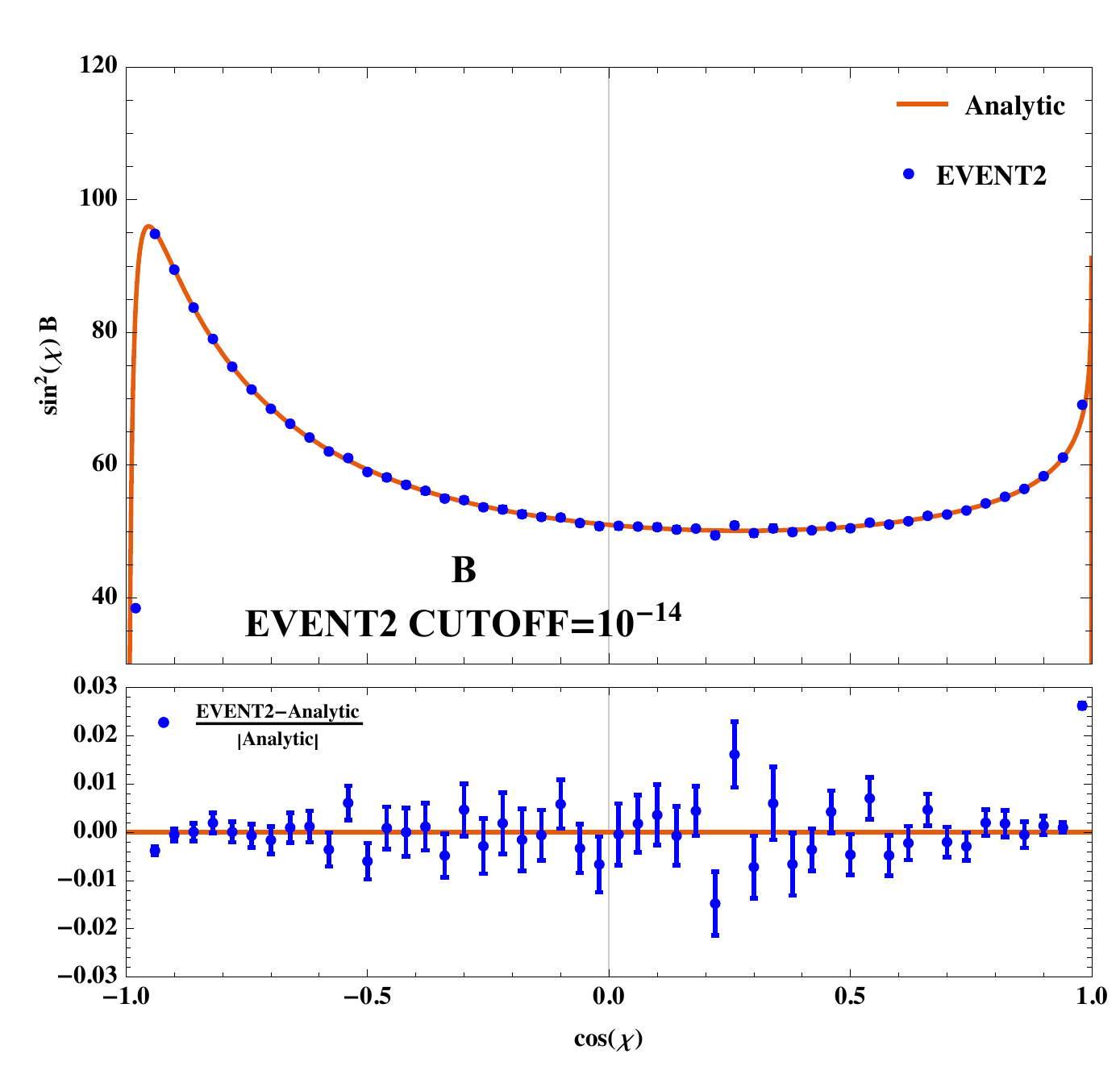}
  \caption{Analytical results for $\sin^2(\chi) B$ are compared with
    numerical results from
    \textsc{Event2}~\cite{Catani:1996jh,Catani:1996vz}.
   The \textsc{Event2} prediction is obtained after sampling over 10
   billion points, with the internal \texttt{CUTOFF} set to $10^{-14}$.
   Error bars represent \textsc{Event2} statistical uncertainties.}
  \label{fig:Btot}
\end{figure}

We have performed a number of checks on the results. First, the individual
virtual and real corrections are IR divergent, but the divergent terms cancel
after summing virtual and real, as required for any IR-safe observable. 
Second, in Fig.~\ref{fig:Btot} we compare our analytical results with
numerical predictions from \textsc{Event2}, which is based on
the dipole subtraction method~\cite{Catani:1996jh,Catani:1996vz}.
We find excellent agreement with \textsc{Event2} over a large range;
the apparent discrepancy in the rightmost bin is mainly due to the
finite bin width used in \textsc{Event2}.
The $z\to 0$ and $z\to 1$ limits of the
analytical results are in perfect agreement with those predicted
respectively by jet calculus~\cite{Konishi:1978ax,Richards:1983sr}
and soft-gluon resummation~\cite{deFlorian:2004mp,MoultZhu2018},
as we discuss in the next section.
%


\noindent\textit{Discussion.} It is interesting to study the end-point
asymptotic limits of the EEC, which provide useful information for
resummation and for constructing more accurate parton showers.
Expanding our results in the $z \to 0$ limit gives
\begin{align}
B(z) = \; & 
C_F \left\{\frac{1}{z} \left[ \log (z) \left(-\frac{107
            C_A}{120}+\frac{25 C_F}{32}+\frac{53 N_f T_f}{240}\right)
\right.\right.
\nn\\
&
\left.\left.
 +C_A \left(-\frac{25
  \zeta_2}{12}+\frac{\zeta_3}{2}+\frac{17683}{2700}\right)
\right.\right.
\nn\\
&
\left.\left.
+C_F \left(\frac{43
  \zeta_2}{12}-\zeta_3-\frac{8263}{1728}\right)-\frac{4913 N_f
  T_f}{3600}\right]
\right.
\nn\\
&
\left.
+\log (z) \left[C_A \left(\frac{33
  \zeta_2}{2}-\frac{703439}{25200}\right)
\right.\right.
\nn\\
&
\left.\left.
+C_F
  \left(\frac{42109}{1200}-21 \zeta_2\right)
+ N_f T_f \left(\frac{86501}{12600}-4 \zeta_2\right)\right]
\right.
\nn\\
&
\left.
+C_A \left(\frac{213 \zeta_2}{5}-\frac{101
  \zeta_3}{2}-\frac{26986007}{5292000}\right)
\right.
\nn\\
&
\left.
+C_F \left(-\frac{1541 \zeta_2}{30}+65
  \zeta_3+\frac{18563}{2700}\right)
\right.
\nn\\
&
\left.
+ N_f T_f \left(-\frac{46 \zeta_2}{3}+12
  \zeta_3+\frac{2987627}{330750}\right)\right\} + {\cal O}(z) \,,
  \label{eq:zto0}
\end{align}
where we have expanded through ${\cal O}(z^0)$.
Note that individual terms in Eq.~\eqref{eq:lc} are far more
singular as $z\to0$ than is the total~\eqref{eq:zto0}.
The EEC in the $z\to0$ limit is dominated by collinear splitting.
The leading-logarithmic term
$\log(z)/z$ has been predicted~\cite{Konishi:1978ax,Richards:1983sr}
using jet calculus~\cite{Konishi:1978yx,Konishi:1978ax}.
The result is expressed as a product
of two $2\times2$ (quark-gluon) anomalous dimension matrices for twist 2, 
spin 3 operators, plus a contribution due to the running coupling. It
agrees fully with the coefficient of $\log(z)/z$ in Eq.~\eqref{eq:zto0}.

In the back-to-back limit, $z \to 1$, we find that the expansion 
of $B(z)$ to next-to-leading power reads
\begin{align}
B(z) = \;& 
C_F \left\{
\frac{1}{1-z} \left[
+\frac{1}{2} C_F \log ^3(1-z)
\right.\right.
\nn\\
& \left.\left.
+ \log ^2(1-z) \left(\frac{11 C_A}{12}+\frac{9
  C_F}{4}-\frac{N_f T_f}{3}\right)
\right.\right.
\nn\\
& \left.\left.
+ \log (1-z) \left(C_A \left(\frac{\zeta_2}{2}-\frac{35}{72}\right)+C_F
  \left(\zeta_2+\frac{17}{4}\right)
\right.\right.\right.
\nn\\
& \left.\left.\left.
+\frac{N_f T_f}{18}\right)
+ C_A \left(\frac{11 \zeta_2}{4}+\frac{3
  \zeta_3}{2}-\frac{35}{16}\right)
\right.\right.
\nn\\
& \left.\left.
+C_F \left(3 \zeta_2-\zeta_3+\frac{45}{16}\right)+ N_f T_f \left(\frac{3}{4}-\zeta_2\right)
\right]
\right.
\nn\\
& \left.
+ \left(\frac{C_A}{2}+C_F\right) \log ^3(1-z)
\right.
\nn\\
& \left.
+ \log ^2(1-z) \left(\frac{27 C_A}{8}+\frac{13 C_F}{2}-\frac{N_f
  T_f}{2}\right)
\right.
\nn\\
& \left.
+ \log (1-z) \left[C_A \left(22 \zeta_2-\frac{2011}{72}\right)
\right.\right.
\nn\\
& \left.\left.
+C_F (47-19 \zeta_2)+ N_f T_f \left(\frac{361}{36}-4
  \zeta_2\right)\right]
\right.
\nn\\
& \left.
+ 
C_A \left(\frac{6347 \zeta_2}{80}-21 \zeta_2 \log (2)-\frac{137
  \zeta_3}{4}-\frac{3305}{72}\right)
\right.
\nn\\
& \left.
+C_F \left(-\frac{1727 \zeta_2}{20}+42 \zeta_2 \log (2)+\frac{121
  \zeta_3}{2}+\frac{3437}{96}\right)
\right.
\nn\\
& \left.
+ N_f T_f \left(-\frac{1747 \zeta_2}{120}+12 \zeta_3+\frac{2099}{144}\right)
\right\} + {\cal O}(1-z) \,.
  \label{eq:zto1}
\end{align}
All the terms enhanced by $(1-z)^{-1}$ were predicted
previously~\cite{deFlorian:2004mp}, in full agreement with Eq.~\eqref{eq:zto1}.
The next-to-leading power terms are new.  They will provide useful information
for resumming large Sudakov logarithms beyond leading
power~\cite{Laenen:2008ux,Laenen:2010uz,Bonocore:2016awd,%
Moult:2016fqy,Boughezal:2016zws,Feige:2017zci,Moult:2017rpl,
Moult:2017jsg,Balitsky:2017flc,Beneke:2017ztn}.
We note the appearance of $\zeta_2 \log(2)$ in the constant term
at next-to-leading power, which originates solely from $B_{\text{nlc}}$. 

\noindent\textit{Summary.} 
We have presented the analytical result for the EEC in QCD at NLO.
Our calculation was enabled by using the IBP equations in a novel way.
The final result turns out to be rather simple; only 11 transcendental
functions are required to describe the QCD results, and these functions
are no more complicated than the ones in the
${\cal N}=4$ SYM result~\cite{Belitsky:2013ofa}.  In contrast,
the polynomial prefactors are of considerably higher degree for QCD.
We have checked our results against \textsc{Event2} numerically and
found full agreement.  We have also expanded the EEC to next-to-leading
power in the collinear and back-to-back limits.
The simplicity of the full NLO result provides encouragement for trying
to compute the EEC at NNLO analytically.
It will also be interesting to apply our method to other event-shape
variables, such as the $C$ parameter (which does appear to require
elliptic functions, even at LO)~\cite{Ellis:1980wv}.

\begin{acknowledgments}
We thank Marc Schreiber for extensive contributions at the beginning
of this project, Alexander Smirnov for helpful instruction on the use of
\textsc{Fire} 5, Oleksandr Gituliar and Vitaly Magerya for their explanations of how to use \textsc{Fuchsia}, and Stefan H\"{o}che for useful discussions.
The work of M.X.L., V.S., T.Z.Y., and H.X.Z.~was supported in part by the
National Science Foundation of China
(11135006, 11275168, 11422544, 11375151, 11535002)
and the Zhejiang University Fundamental Research Funds for the
Central Universities (2017QNA3007). H.X.Z.~was also supported by the 
Thousand Youth Program of China.
The research of L.D.~was supported by the US Department of Energy under
contract DE--AC02--76SF00515.
\end{acknowledgments}

\bibliography{EEC_NLO.bib}{}
\bibliographystyle{apsrev4-1}


\newpage

\onecolumngrid
\newpage
\appendix

\section*{Supplemental material}

In this supplemental material, we provide detailed analytic formulae for the
remaining coefficients in the color decomposition of the EEC at NLO,
as well as for an identical-quark piece that cannot be isolated by color alone.
We describe the $z\to-\infty$ limit of the various color components,
and provide the $z\to0$, 1 and $-\infty$ limits of the identical-quark
piece.  The $z\to0$ limit of this piece can be interpreted using
jet calculus at next-to-leading logarithm.
Finally, we validate our analytic results for each color component
numerically against \textsc{Event2}.


\subsection{Analytical results for remaining color coefficients}
\label{sec:analyt-results-rema}

\begin{figure}
  \begin{subfigure}[b]{0.4\textwidth}
    \includegraphics[width=\textwidth]{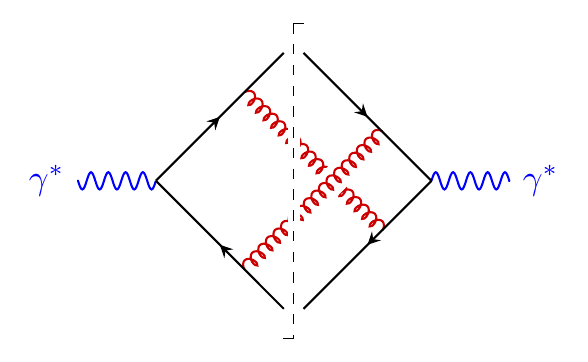}
    \caption{$qqgg$}
    \label{fig:gg}
  \end{subfigure}
  \begin{subfigure}[b]{0.4\textwidth}
    \includegraphics[width=\textwidth]{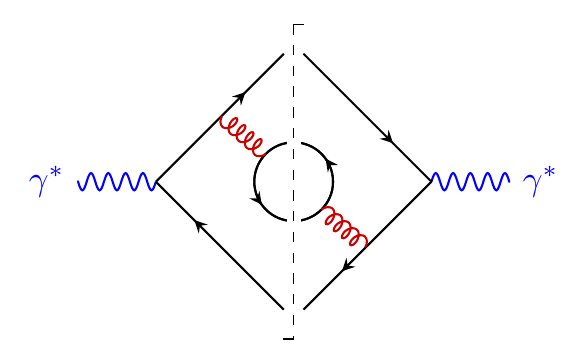}
    \caption{$q\bar{q}q'\bar{q}'$}
    \label{fig:flavor}
  \end{subfigure}
  \begin{subfigure}[b]{0.4\textwidth}
    \includegraphics[width=\textwidth]{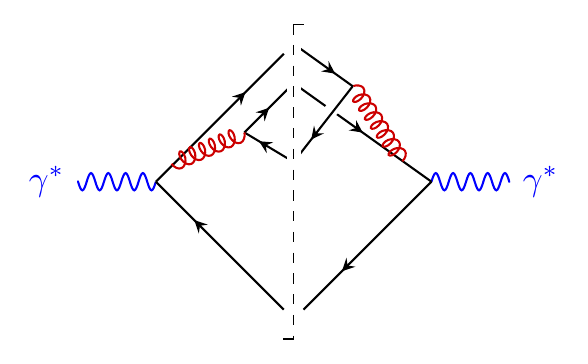}
    \caption{$q\bar{q}q\bar{q}$}
    \label{fig:qq}
  \end{subfigure}
  \begin{subfigure}[b]{0.4\textwidth}
    \includegraphics[width=\textwidth]{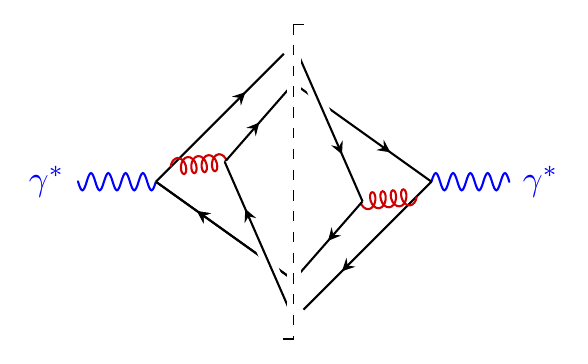}
    \caption{$q\bar{q}q\bar{q}$}
    \label{fig:qq0}
  \end{subfigure}
\caption{Representative cut diagrams for real corrections to the EEC at NLO.}
\end{figure}

The most complicated part of this calculation involves the real
corrections. We show representative cut diagrams for real
corrections with final states $q\bar{q}gg$~(Fig.~\ref{fig:gg}),
$q\bar{q}q'\bar{q}'$~(Fig.~\ref{fig:flavor}), and
$q\bar{q}q\bar{q}$~(Fig.~\ref{fig:qq} and \ref{fig:qq0}). Note that
the class of diagrams in Fig.~\ref{fig:qq0} vanishes for a virtual photon
thanks to Furry's theorem, because there are three vector couplings
attached to each fermion ``loop'' and the measurement function is
invariant under charge conjugation~\cite{Ellis:1980wv}. (At the $Z$ pole,
the purely axial-vector contributions represented by Fig.~\ref{fig:qq0}
vanish for the first two generations in the massless quark limit,
while the third generation contribution is tiny~\cite{Hagiwara:1990dx}.)

The subleading color coefficient $B_{\text{nlc}}$
appearing in Eq.~\eqref{eq:colordec} is given by
\begin{subequations}
\begin{align}
  B_{\text{nlc}} = \; 
& +\frac{57600 z^7-115200 z^6+75748 z^5-17359 z^4+902 z^3+14966 z^2-27552 z+9320}{720 (1-z) z^4}
\nn \\
& -\frac{-115200 z^9+316800 z^8-321680 z^7+147846 z^6-31035 z^5+3225 z^4-3571 z^3+11322 z^2-12412 z+4880}{360 (1-z) z^5} g^{(1)}_1
\nn \\
& -\frac{230400 z^8-518400 z^7+412960 z^6-138600 z^5+18696 z^4-742 z^3+10971 z^2-25029 z+11424}{720 (1-z) z^4} g^{(1)}_2
\nn \\
& +\frac{-91 z^7+235 z^6-184 z^5+15 z^4-140 z^3+721 z^2-760 z+314}{120 (1-z) z^5} g^{(2)}_1
\nn \\
& +\frac{-19200 z^8+28800 z^7-14680 z^6+2660 z^5-340 z^4-40 z^3+315 z^2-1431 z+952}{60 z^5} g^{(2)}_2
\nn \\
& +\frac{960 z^4-160 z^3+992 z^2+547 z+1435}{480 z^{7/2}} g^{(2)}_3
 -\frac{-120 z^6+120 z^5-130 z^4-585 z^3+2647 z^2-3143 z+1266}{60 (1-z) z^5} g^{(2)}_4
\nn \\
& +\frac{640 z^6-1920 z^5+2196 z^4-1196 z^3+318 z^2-42 z+3}{4 (1-z) z} g^{(3)}_1
 +\frac{2 z^7-3 z^6+3 z^5-z^4-z^3+9 z^2-9 z+1}{12 (1-z) z^5} g^{(3)}_2
\nn \\
& -\frac{(1-2 z) \left(z^2-z+1\right)}{2 (1-z) z} g^{(3)}_4
 -\frac{2 z^5-z^4+2 z^3+z^2+3}{4 z^4} g^{(3)}_5 \,,
  \label{eq:nlc}
\end{align}
\end{subequations}
where we have introduced two more weight 3 transcendental functions,
\begin{subequations}
\begin{align}
  g_4^{(3)} = \; & \text{Li}_3\left(-\frac{z}{1-z}\right)
                - 3 \, \zeta_2 \log(z) + 8 \, \zeta_3 \,,
\nn\\
g_5^{(3)} = \; & 
- 8 \left[ \text{Li}_3\left(-\frac{\sqrt{z}}{1-\sqrt{z}}\right)
+ \text{Li}_3\left(\frac{\sqrt{z}}{1+\sqrt{z}}\right) \right]
+ 2 \text{Li}_3\left(-\frac{z}{1-z}\right)
+ 4 \zeta_2 \log (1-z)+\log \left(\frac{1-z}{z}\right)
     \log^2\left(\frac{1+\sqrt{z}}{1-\sqrt{z}}\right) \,.
  \label{eq:moreGfunc}
\end{align}
\end{subequations}
Note that $g_5^{(3)}$ is even under $\sqrt{z} \to - \sqrt{z}$.

The contribution to Eq.~\eqref{eq:colordec} that is proportional
to the number of light quark flavors, $N_f$, is given by
\begin{subequations}
\begin{align}
  \label{eq:Bnf}
  B_{N_f} = \;
& -\frac{7200 z^7-14400 z^6+8852 z^5-1568 z^4+48 z^3+1825 z^2-4115 z+2050}{144 (1-z) z^4}
\nn \\
& -\frac{72000 z^9-198000 z^8+193040 z^7-77700 z^6+10960 z^5-100 z^4-489 z^3+3269 z^2-4801 z+1801}{360 (1-z) z^5} g^{(1)}_1
\nn \\
& +\frac{36000 z^8-81000 z^7+60520 z^6-16650 z^5+1190 z^4+10 z^3+428 z^2-939 z+561}{180 (1-z) z^4} g^{(1)}_2
\nn \\
& +\frac{-z^7-4 z^3+18 z^2-24 z+9}{6 (1-z) z^5} g^{(2)}_1
 -\frac{-12000 z^8+18000 z^7-7840 z^6+920 z^5+72 z^2-222 z+187}{60 z^5} g^{(2)}_2
\nn \\
& +\frac{1-3 z}{48 z^{7/2}} g^{(2)}_3
 +\frac{8 z^3-66 z^2+71 z+7}{60 (1-z) z^5} g^{(2)}_4
 +2 \left(50 z^4-100 z^3+66 z^2-16 z+1\right) g^{(3)}_1 \,.
\end{align}
\end{subequations}
It is straightforward to take the limits of $B_{\text{lc}}$,
$B_{\text{nlc}}$ and $B_{N_f}$ as $z\to0$ and $z\to1$, in order
to obtain Eqs.~\eqref{eq:zto0} and \eqref{eq:zto1}, as well as
further subleading powers if desired.

Another limit that we can study with the exact NLO result is
$z\to \infty$, where the limit can be taken in any radial
direction on the complex plane.
Although this limit requires an analytic continuation
out of the physical region, the analytic properties may still prove useful
for understanding the EEC as a limit of a four-point
correlator~\cite{Hofman:2008ar,%
Belitsky:2013ofa,Belitsky:2013xxa,Belitsky:2013bja},
and perhaps for constructing it at higher perturbative orders.
In fact, we find that the LO result and the NLO coefficients
are quite suppressed in this limit.
In the NLO case, the suppressed behavior requires cancellations between
many different terms in the exact result.
The leading-power terms as $z \to \infty$ have the form,
\begin{align}
  \label{eq:Aztoinfty}
 A(z) = \; &
\frac{C_F}{z^3} \left[ 2 \, \log(-z) - \frac{9}{2} \right]
+ {\cal O}(1/z^4) \,, \\
  \label{eq:Blctoinfty}
 B_{\text{lc}}(z) = \; &
\frac{1}{z^3} \left[
   \left( 4 \, \zeta_2 + \frac{4699}{288} \right) \log(-z)
   - 8 \, \zeta_3 + \frac{991}{84} \, \zeta_2 - \frac{85595}{1728} \right]
\nn\\
 \; & + \frac{i \, \text{sign}(\text{Im}(z))}{z^3} \left[  \frac{11}{8}\, \zeta_2 \sqrt{-z} + \pi 
  \left( - \frac{1459}{140} \log(-z) + \frac{466259}{19600} \right)  \right] 
+ {\cal O}(1/z^{7/2}) \,, \\
  \label{eq:Bnlctoinfty}
 B_{\text{nlc}}(z) = \; &
\frac{1}{z^3} \left[
   \left( \frac{3}{2} \, \zeta_2 + \frac{473}{72} \right) \log(-z)
   - \frac{9}{2} \, \zeta_3 + \frac{521}{70} \, \zeta_2
   - \frac{32713}{1728} \right]
\nn\\
\; & + \frac{i \, \text{sign}(\text{Im}(z))}{z^3} \left[ - \frac{2059}{560} \, \zeta_2 \sqrt{-z} + \pi
   \left( - \frac{2407}{420} \log(-z) + \frac{3}{2}\,\zeta_2 + \frac{20518}{1225} \right)\right]
+ {\cal O}(1/z^{7/2}) \,, \\
  \label{eq:Bnftoinfty}
 B_{N_f}(z) = \; &
\frac{1}{z^3} \left[ - \frac{133}{36} \log(-z)
   - \frac{404}{105} \, \zeta_2  + \frac{51}{4} \right]
\nn\\
\; &
 + \frac{i \, \text{sign}(\text{Im}(z))}{z^3} \left[ - \frac{3}{8}\, \zeta_2 \sqrt{-z} + \pi 
\left( \frac{26}{21} \log(-z) - \frac{196003}{88200} \right) \right] 
+ {\cal O}(1/z^{7/2}) \,.
\end{align}
Here the branch cut is chosen along the negative real axis for both
$\log(-z)$ and $\sqrt{-z}$. 

In the the remainder of this subsection, we discuss the contribution
from identical-quark exchange terms of the type shown in Fig.~\ref{fig:qq}.
In the real corrections, both the $q\bar{q}gg$ final state
and the $q\bar{q}q\bar{q}$ final state with identical quarks
contribute to the subleading color coefficient.
We can write the real corrections to $B_{\text{nlc}}$ as
\begin{subequations}
\begin{align}
  \label{eq:nlcdecomp}
  B^R_{\text{nlc}} = B^R_{g} + B_{qq_\text{int}} \,.
\end{align}
\end{subequations}
There are no virtual corrections to the identical-quark interference terms
at this order. Hence $B_{qq_\text{int}}$ is by itself IR finite and
gauge invariant, although it is just one piece of the
color component $B_{\text{nlc}}$.
It has the following form,
\begin{subequations}
\begin{align}
B_{qq_\text{int}} = 
& -\frac{-79200 z^7+158400 z^6-104356 z^5+24394 z^4-1028 z^3-3001 z^2-7735 z+13696}{1440 (1-z) z^4}
\nn \\
& +\frac{-79200 z^8+138600 z^7-83680 z^6+20096 z^5-2880 z^4+28 z^3+1146 z^2-2812 z+1637}{360 z^5} g^{(1)}_1
\nn \\
& +\frac{-79200 z^8+178200 z^7-143080 z^6+48870 z^5-7253 z^4+231 z^3+1448 z^2-4239 z+5358}{360 (1-z) z^4} g^{(1)}_2
\nn \\
& -\frac{-202 z^6+220 z^5-135 z^4-100 z^3+520 z^2-748 z+325}{240 z^5} g^{(2)}_1
\nn \\
& -\frac{ \left(13200 z^8-19800 z^7+10280 z^6-1930 z^5+345 z^4+75 z^3+41 z^2+225 z+113\right)}{60 z^5} g^{(2)}_2
\nn \\
& +\frac{480 z^5-80 z^4+496 z^3+375 z^2+320 z+240}{240 z^{9/2}} g^{(2)}_3
\nn \\
& +\frac{240 z^6-240 z^5+260 z^4-592 z^3+840 z^2-969 z+551}{120 (1-z) z^5} g^{(2)}_4
 +\frac{15 z^2-68 z+96}{4 \sqrt{1-z} z^{9/2}} g^{(2)}_5
\nn \\
& +\frac{-440 z^5+880 z^4-636 z^3+199 z^2-31 z+3}{4 z} g^{(3)}_1
 -\frac{(1-2 z)(z^2-z+1)}{2 (1-z) z}
   \left(\frac{g^{(3)}_2}{6}+g^{(3)}_4+\frac{g^{(3)}_5}{2}\right) 
\nn \\
& +\frac{z^2-11 z+22}{8 z^5}
    \left(-\frac{1}{2} g^{(3)}_2+g^{(3)}_5-\frac{g^{(3)}_6}{2}+g^{(3)}_7\right)
 +\frac{-z^3+12 z^2-33 z+24}{24 (1-z) z^5}
   \left(-3 g^{(3)}_5-33 g^{(3)}_8+g^{(3)}_9\right)\,.
\label{eq:qqint}
\end{align}
\end{subequations}
The one additional weight 2 and four additional weight 3
transcendental functions introduced to describe $B_{qq_\text{int}}$ are defined as
\begin{subequations}
\begin{align}
  \label{eq:evenMoreGfunc}
  g_{5}^{(2)} = \; & \frac{1}{i} \left[ \text{Li}_2(i r) - \text{Li}_2(-i r)
        - \log(r) \log\left(\frac{1+i r}{1-i r}\right) \right] \,,
\nn\\
  g_{6}^{(3)} = \; & \log ^3(1-z)-15 \zeta_2 \log (1-z) \,,
\nn\\
  g_{7}^{(3)} = \; & \log (1-z) \left(\text{Li}_2(z)+\log (1-z) \log (z)-\frac{15 \zeta_2}{2}\right) \,,
\nn\\
  g_{8}^{(3)} = \; & \zeta_3 \,,
\nn\\
  g_{9}^{(3)} = \; & -12 \left[
 -\text{Li}_3\left(\frac{1}{2} (1-i r)\right)
 -\text{Li}_3\left(\frac{1}{2} (1+i r)\right)
 +\text{Li}_3(-i r)+\text{Li}_3(i r)
 +\text{Li}_3\left(-\frac{2 r}{i-r}\right)
 +\text{Li}_3\left(\frac{2 r}{i+r}\right) - \zeta_3 \right]
\nn\\
&
+ 3 \, \text{Li}_3\left(-\frac{z}{1-z}\right)
 + 2 \left[ \log^3\left(\frac{1}{2} (1-i r)\right)
          +\log^3\left(\frac{1}{2} (1+i r)\right) \right]
 - 3 \, (2 \log (i r)-i \pi ) \, \log^2 \left(\frac{1-i r}{1+i r}\right)
\nn\\
&
-\pi ^2 \left(\log \left(\frac{1}{2} (1-i r)\right)+\log \left(\frac{1}{2} (1+i r)\right)\right) \,,
\end{align}
\end{subequations}
where $r = \sqrt{z}/\sqrt{1-z}$.  The function $g^{(2)}_5$ is
also known as the Bloch-Wigner function~\cite{BlochWigner}.
It is single-valued (real analytic) in $z=ir$, $\bar{z}=-ir$;
here we only need it on the imaginary $z$ axis.
Although $g^{(2)}_5$ and $g^{(3)}_9$
contain logarithms and classical polylogarithms with complex arguments,
they are real for $z \in (0,1)$.  While $g^{(3)}_9$ is even under $r \to -r$,
$g^{(2)}_5$ is odd under this transformation.  On the other hand,
the square roots in the prefactor for $g^{(2)}_5$ in Eq.~\eqref{eq:qqint}
ensure that only integer powers of $z$ and $1-z$ appear in the expansions
around these limits.

Interestingly, the identical-quark interference terms are more complicated
than the final results for $B_{\text{nlc}}$
in Eq.~\eqref{eq:nlc}, in the sense that five more
transcendental functions are needed to fully describe them,
including $g^{(2)}_5$ and $g^{(3)}_9$ with their complex polylogarithmic arguments.
In the full NLO result, the coefficients of these functions cancel
between the $q\bar{q}gg$ cuts and the $q\bar{q}q\bar{q}$ cuts.

It also worth pointing out that while the identical-quark interference
contribution is IR finite by itself, it does have a leading-power
singularity (but not leading-log) in both the collinear and back-to-back
regions.  Expanding the result~\eqref{eq:qqint}, we find the asymptotic
$z \to 0$ limit to next-to-leading power (NLP) is given by
\begin{subequations}
  \begin{align}
B_{qq_\text{int}} \overset{z \to 0}{=} \; &
\frac{1}{z} \left( \frac{\zeta_3}{2}-\frac{43 \zeta_2}{24}+\frac{8011}{3456}\right)+\left(\frac{25 \zeta_2}{2}-\frac{147893}{7200}\right) \log (z)-\frac{77 \zeta_3}{2}+\frac{463 \zeta_2}{15}-\frac{652897}{144000} + {\cal O}(z) \,,
\label{eq:qqintzto0}    
  \end{align}
\end{subequations}

Remarkably, the coefficient of $1/z$ in Eq.~\eqref{eq:qqintzto0}
can be reproduced using the jet
calculus approach at next-to-leading-logarithm~\cite{Kalinowski:1980wea}.
While the leading-log $\log(z)/z$ contribution in the full NLO QCD result
comes from iterating two $1\to2$ splittings, $q\to qg$ and $g\to gg$,
and accounting for running-coupling effects~\cite{Richards:1983sr},
the next-to-leading logarithms also involve the $1\to3$ splitting process.
The identical-quark exchange terms in the triple-collinear splitting
process $q \rightarrow q q \bar{q}$ are somewhat unique in not requiring
the subtraction of any iterated $1\to2$ splittings.
Assigning longitudinal momentum fractions $x_1$, $x_2$ and $x_3=1-x_1-x_2$
to the three quarks, these exchange terms can be described by the function
\begin{eqnarray}
E(x_1,x_2) &=& 2 \left[ - 2 + \frac{1+x_1}{1-x_2} + \frac{1+x_2}{1-x_1}
      - \frac{2x_1}{(1-x_2)^2} - \frac{2x_2}{(1-x_1)^2} \right. \nn\\
&&\null
\left.
 + \left( 2 - \frac{1+x_1}{1-x_2} - \frac{1+x_2}{1-x_1}
        + \frac{2}{(1-x_1)(1-x_2)} \right)
    \log \left(\frac{(1-x_1)(1-x_2)}{1-x_1-x_2}\right) \right]
\label{eq:Ex1x2}
\end{eqnarray}
given in Eq.~(B.7) of Ref.~\cite{Kalinowski:1980wea}.
The relative transverse momenta have already been integrated over
to obtain $E(x_1,x_2)$.  Assuming that all angles between the three
final-state partons are comparable,
we multiply $E(x_1,x_2)$ by $x_1x_2+x_2x_3+x_3x_1$ in order to account
for the energy weighting in the EEC,
and integrate over the remaining longitudinal momentum fractions,
\begin{equation}
\int_0^1 dx_1 \int_0^{1-x_2} dx_2 \, E(x_1,x_2) \, [ x_1x_2 + (x_1+x_2)(1-x_1-x_2) ]
\ =\ - 4 \, \zeta_3 + \frac{43}{3} \, \zeta_2 - \frac{8011}{432} \,.
\end{equation}
Accounting for the constant prefactors in Ref.~\cite{Kalinowski:1980wea},
we reproduce the $1/z$ term in Eq.~\eqref{eq:qqintzto0}.
For other components of $B(z)$,
in which there is also a $\log(z)/z$ term, the $1/z$ term is subleading
and the integration of the analogous functions in
Ref.~\cite{Kalinowski:1980wea} does not
reproduce the correct result, presumably because the assumption of
comparable angles between the three partons fails. It may still be possible
to extract the $1/z$ term in this case by using the full dependence
of the triple-collinear splitting process on the two-particle invariants
$p_i\mcdot p_j$~\cite{Campbell:1997hg,Catani:1998nv}.

In the opposite, back-to-back region, the expansion of $B_{qq_\text{int}}$
to NLP is given by
\begin{align}
  \label{eq:qqintzto1}
  B_{qq_\text{int}} \overset{z \to 1}{=} \; &
\frac{1}{1-z}
 \left( - \frac{\zeta_3}{2} + \frac{3 \zeta_2}{4} - \frac{13}{16}\right)
+ \frac{3}{4} \log^2(1-z)
+ \left(22 \zeta_2-\frac{255}{8}\right) \log(1-z)
\nn\\
&
- \frac{243 \zeta_3}{4} - 21 \zeta_2 \log (2)
+ \frac{17119 \zeta_2}{240} - \frac{1003}{96} + {\cal O}(1-z) \,.
\end{align}

In the limit as $z\to \infty$, the identical-quark exchange terms
have the same falloff as the other NLO color components:
\begin{align}
  \label{eq:Bqqtoinfty}
  B_{qq_\text{int}}(z) = \; &
\frac{1}{z^3} \left[
   \left( \frac{1}{4} \, \zeta_2 - \frac{59}{72} \right) \log(-z)
   - \frac{1}{8} \, \zeta_3 + \frac{3}{2} \, \zeta_2 \log(2)
   + \frac{407}{420} \, \zeta_2 - \frac{871}{432} \right]
\nn\\
\;& 
 - \frac{i \, \text{sign}(\text{Im}(z))}{z^3} \frac{319}{280}\, \zeta_2 \sqrt{-z} 
+ {\cal O}(1/z^{7/2}) \,.
\end{align}
%


\subsection{Validation of individual color components}
\label{sec:furth-valid-results}

\begin{figure}
  \begin{subfigure}[b]{0.32\textwidth}
    \includegraphics[width=\textwidth]{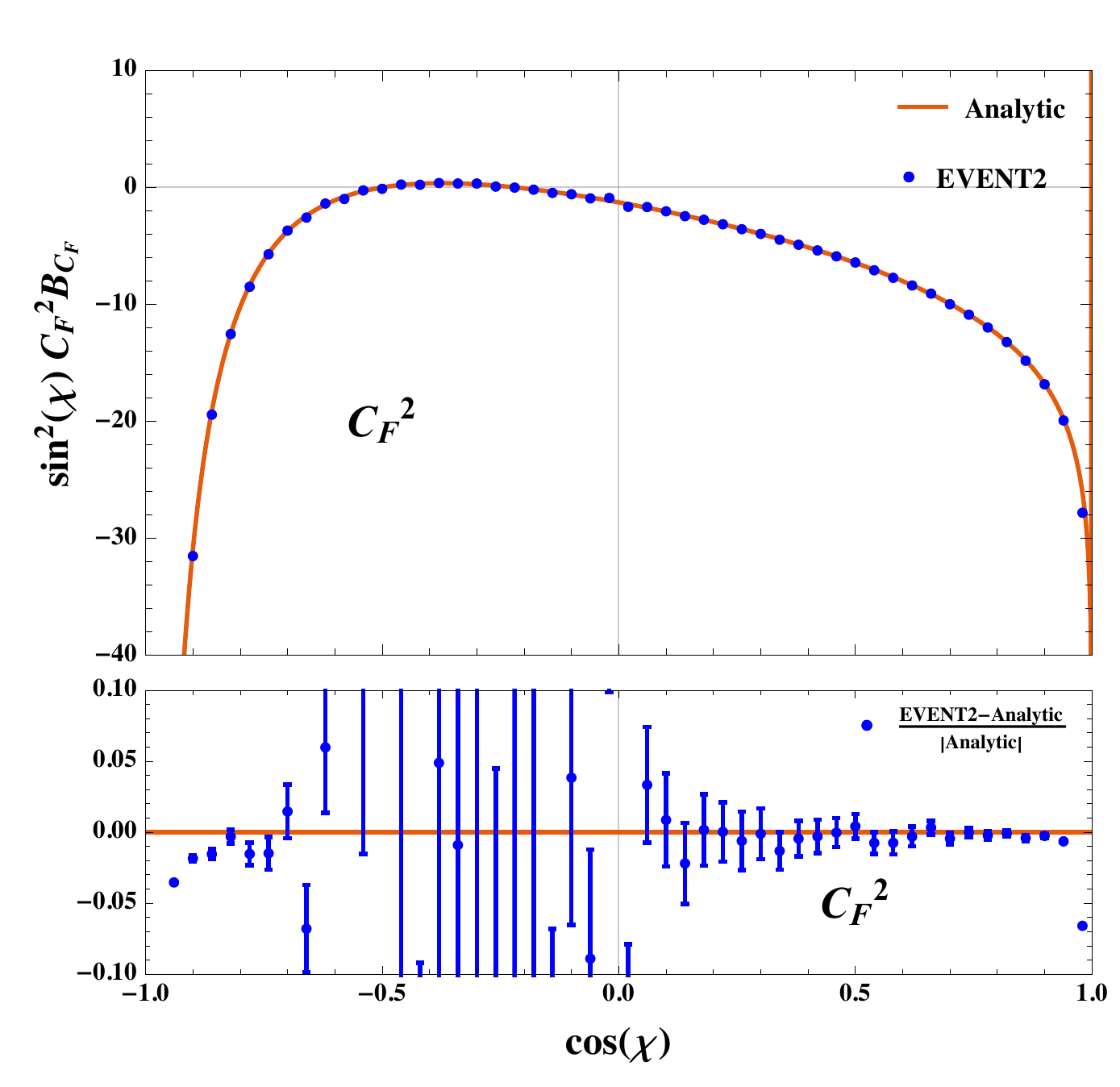}
    \caption{$B_{C_F}$}
    \label{fig:cf}
  \end{subfigure}
  \begin{subfigure}[b]{0.32\textwidth}
    \includegraphics[width=\textwidth]{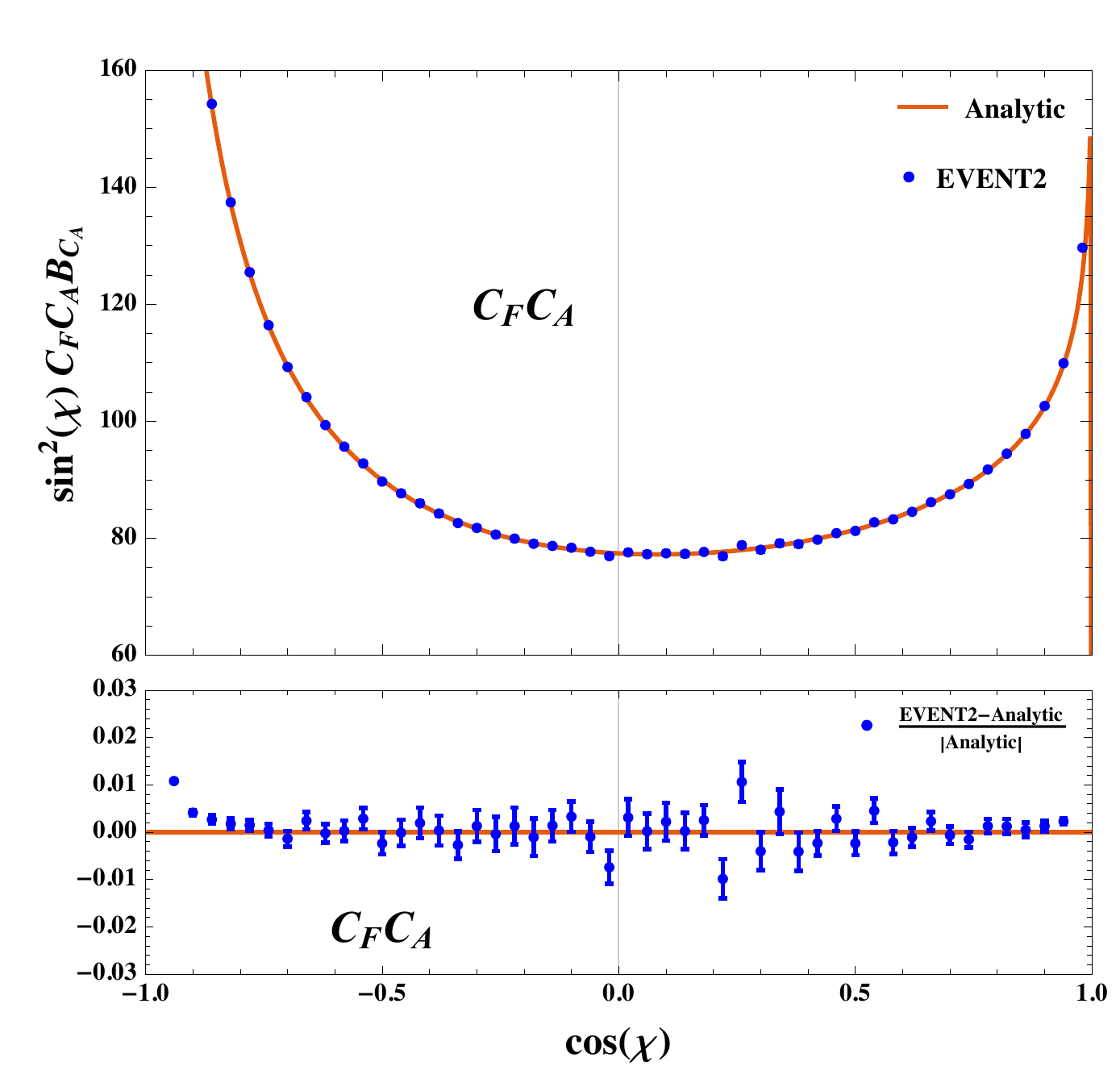}
    \caption{$B_{C_A}$}
    \label{fig:ca}
  \end{subfigure}
  \begin{subfigure}[b]{0.32\textwidth}
    \includegraphics[width=\textwidth]{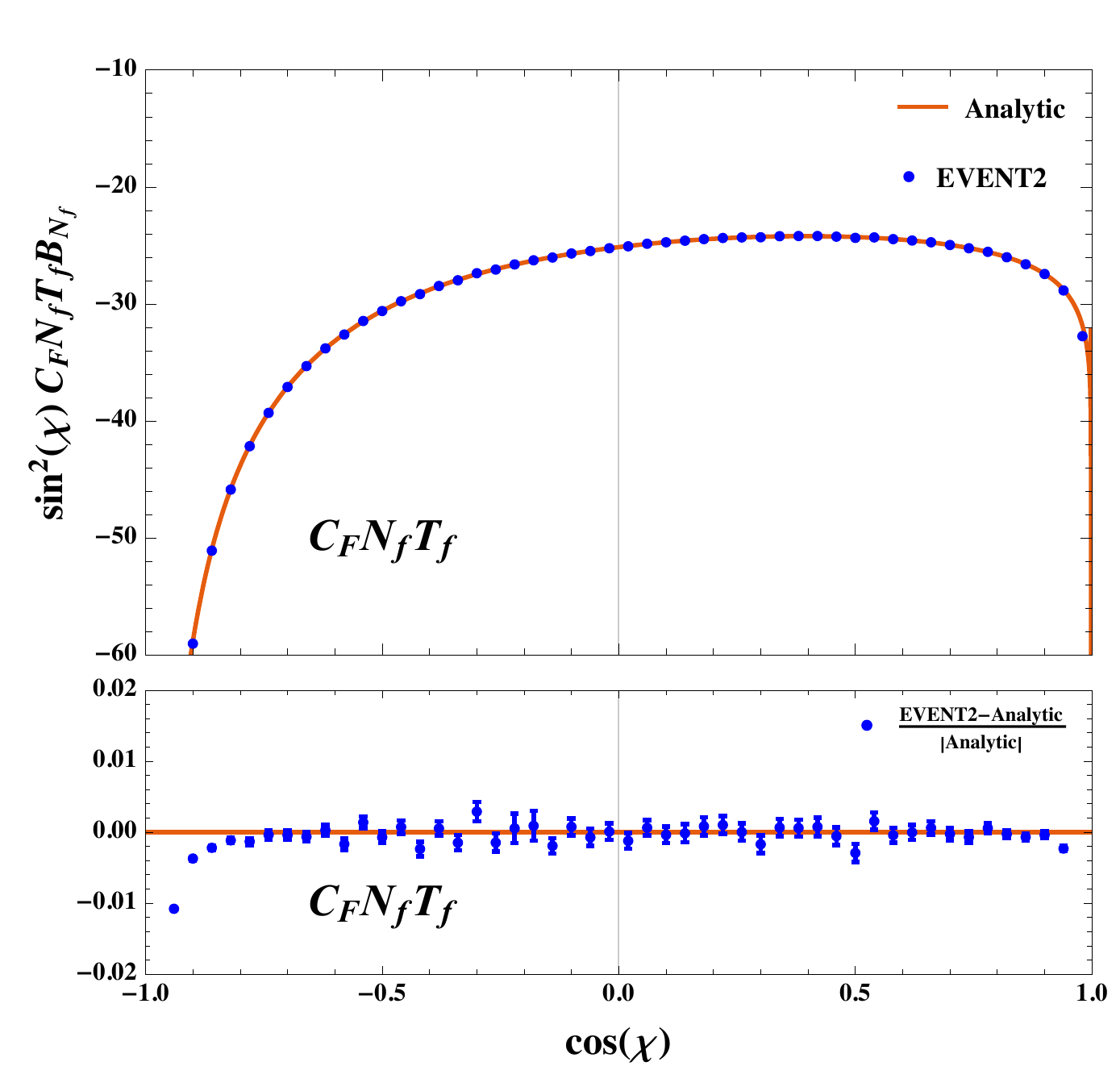}
    \caption{$B_{N_f}$}
    \label{fig:nf}
  \end{subfigure}
   \caption{Comparison with \textsc{Event2} for the $B_{C_F}$, $B_{C_A}$
 and $B_{N_f}$ contributions.}
 \end{figure}

\begin{figure}[ht]
  \centering
  \includegraphics[width=0.5\textwidth]{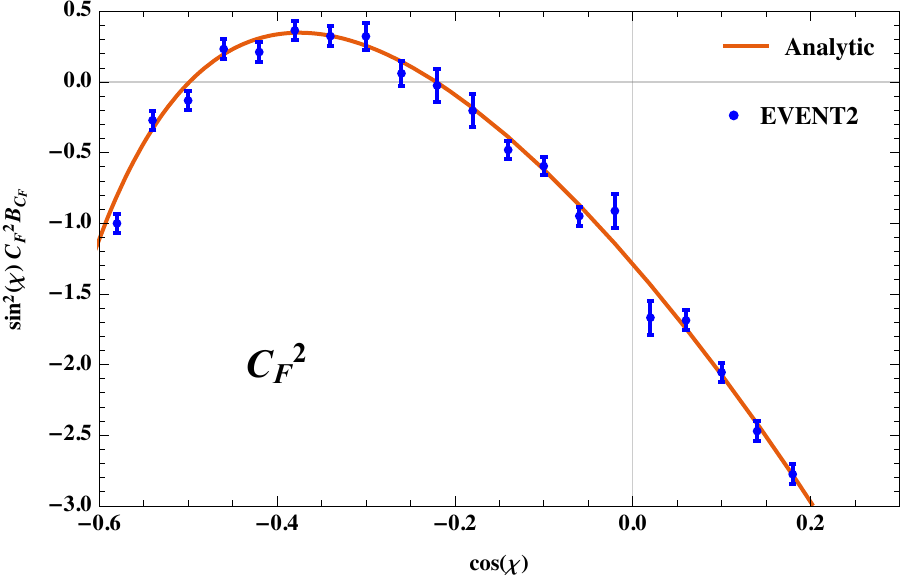}
  \caption{Comparison with \textsc{Event2} for the $C_F^2$ coefficient
  in the region where $B_{C_F}$ is close to zero.}
  \label{fig:cfZoomIn}
\end{figure}

\begin{figure}[h]
  \centering
  \includegraphics[width=0.5\textwidth]{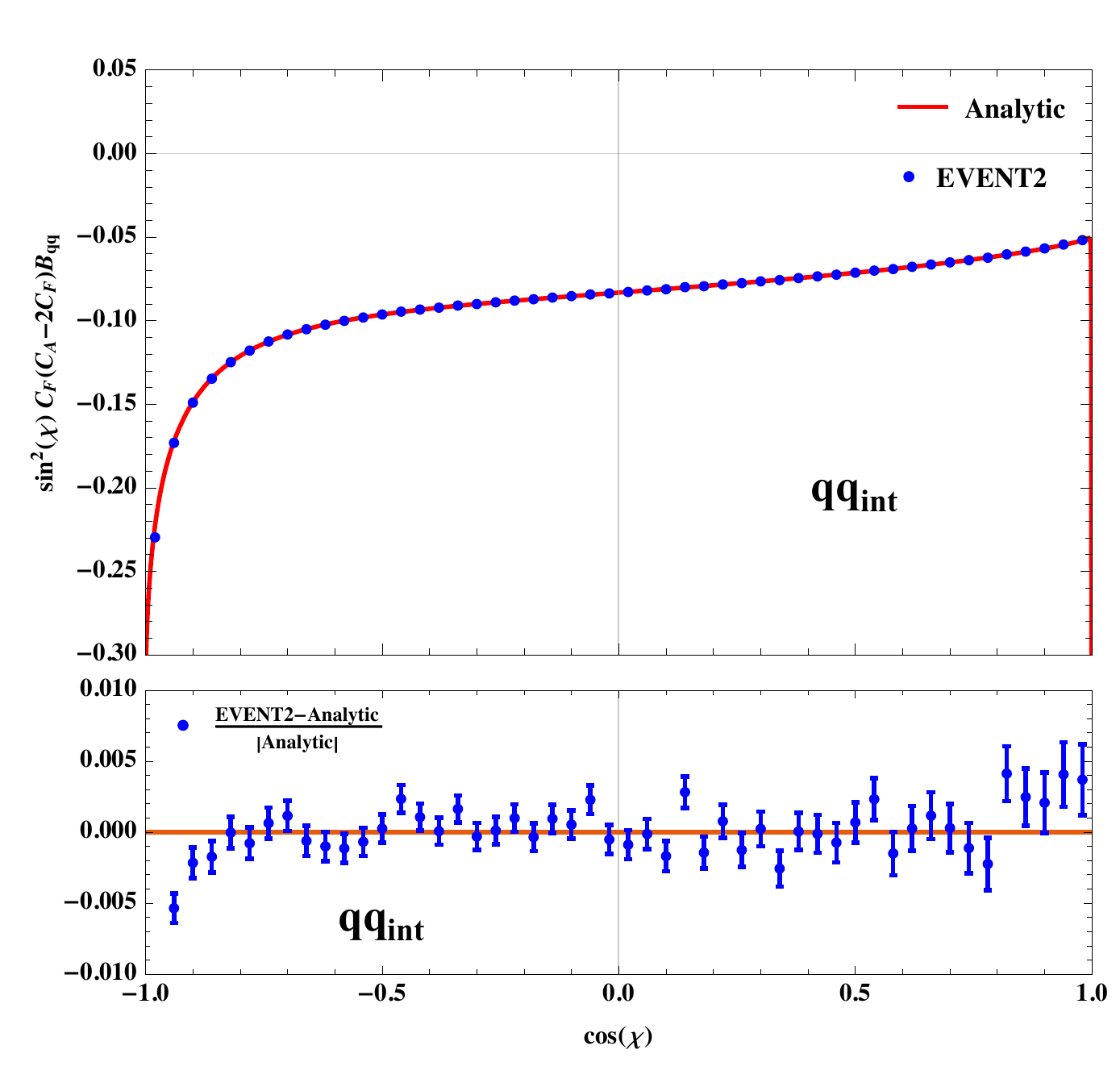}
  \caption{Comparison with \textsc{Event2} for the identical-quark
 interference terms.}
  \label{fig:qqint}
\end{figure}

For validating the color components of the NLO result, we use a slightly
different color basis.  We decompose the NLO coefficient $B$ in terms of
the following color structures:
\begin{align}
  \label{eq:color}
  B = C_F^2 B_{C_F} + C_F C_A B_{C_A} + C_F N_f T_f B_{N_f} \,.
\end{align}
Each of these coefficients can be computed separately using \textsc{Event2}.
In Figs.~\ref{fig:cf},~\ref{fig:ca},~\ref{fig:nf} we compare our analytic
results with numerical results from \textsc{Event2}.
We sample \textsc{Event2} at over 10 billion points, and set its internal
\texttt{CUTOFF} to $10^{-14}$.
Overall we find excellent agreement, except perhaps for $B_{C_F}$ in the
range $-0.5 < \cos\chi < 0$, where large statistical fluctuations are seen
in the bottom panel of Fig~\ref{fig:cf}.  Numerical evaluation of $B_{C_F}$
in this region is challenging, as there are large cancellations between the
virtual and real corrections. A close-up of this region is shown
in Fig.~\ref{fig:cfZoomIn}.  Finally, in Fig.~\ref{fig:qqint} we plot
the identical-quark interference terms, $B_{qq_\text{int}}$.

\end{document}